\documentclass[twocolumn,showpacs,preprintnumbers,amsmath,amssymb,pre]{revtex4}

\usepackage{graphicx}
\usepackage{dcolumn}
\usepackage{bm}
\usepackage{enumerate}
\usepackage{amsmath}
\usepackage{color}

\begin{document}

\title{Subdiffusion equation with Caputo fractional derivative with respect to another function}

\author{Tadeusz Koszto{\l}owicz}
 \email{tadeusz.kosztolowicz@ujk.edu.pl}
 \affiliation{Institute of Physics, Jan Kochanowski University,\\
         Uniwersytecka 7, 25-406 Kielce, Poland}

 \author{Aldona Dutkiewicz}
 \email{szukala@amu.edu.pl}
 \affiliation{Faculty of Mathematics and Computer Science,\\
Adam Mickiewicz University, Uniwersytetu Pozna\'nskiego 4, 61-614 Pozna\'n, Poland}

\date{\today}

\begin{abstract}

We show an application of a subdiffusion equation with Caputo fractional time derivative with respect to another function $g$ to describe subdiffusion in a medium having a structure evolving over time. In this case a continuous transition from subdiffusion to other type of diffusion may occur. The process can be interpreted as ``ordinary'' subdiffusion with fixed subdiffusion parameter (subdiffusion exponent) $\alpha$ in which time scale is changed by the function $g$. As an example, we consider the transition from ``ordinary'' subdiffusion to ultraslow diffusion. The $g$--subdiffusion process generates the additional aging process superimposed on the ``standard'' aging generated by ``ordinary'' subdiffusion. The aging process is analyzed using coefficient of relative aging of $g$--subdiffusion with respect to ``ordinary'' subdiffusion. The method of solving the $g$-subdiffusion equation is also presented. 
\end{abstract}

\maketitle

\section{Introduction\label{secI}}

A type of diffusion is usually defined by time evolution of the Mean Square Displacement (MSD) $\sigma^2(t)$ of a diffusing particle. If $\sigma^2(t)\sim t^\alpha$, we have superdiffusion for $\alpha>1$, normal diffusion for $\alpha=1$ and ``ordinary'' subdiffusion when $0<\alpha<1$. If $\sigma^2(t)\sim \eta(t)$, where $\eta$ is a slowly varying function, we have ultraslow diffusion (slow subdiffusion). A slowly varying function fulfils the condition $\eta(at)/\eta(t)\rightarrow 1$ when $t\rightarrow\infty$ for any $a>0$. In practice, slowly varying function is considered as a combination of logarithmic functions. 
Within the Continuous Time Random Walk (CTRW) model subdiffusion is defined as a process in which a time distribution between particle jumps has a heavy tail which makes the average time infinite, but the jump length distribution has finite moments \cite{bg,mk,mk1,ks,compte,barkai2000,barkai2002}. This process occurs in media, such as gel, where particles diffusion is very hindered \cite{tk2005}. Recently, it has been shown that a membrane which can retain diffusing molecules for a very long time generates subdiffusion in an external medium \cite{kd2021}. Subdiffusion is described by the equation with integral operators with respect to time variable \cite{mk,mk1,ks,compte,barkai2000,barkai2001,barkai2002,schneider}. The operators are usually defined as the Riemann-Liouville fractional time derivative of order $1-\alpha$ or the Caputo fractional time derivative of order $\alpha$. Ultraslow diffusion is an extremely slow process, qualitatively different from ``ordinary'' subdiffusion. It is described by integro--differential equations with the integral operator which is not identified frequently as a fractional time derivative \cite{tk1,tk2015}. This process was observed in diffusion of water in aqueous sucrose glasses \cite{zorbist} and languages dynamics \cite{watanabe}. Superdiffusion is a process in which anomalously long jumps of a particle can be made with a relatively high probability, as in a turbulent medium \cite{bg,mk,mk1} and in motion of endogenous intracellular particles in some pathogens \cite{reverey}. The probability distribution of jump length has a heavy tail while the average waiting time for the particle to jump is finite. The CTRW model provides superdiffusion equation with the fractional Riesz derivative with respect to a spatial variable \cite{mk,mk1,ks,compte}. 

The parameter $\alpha$ depends on the medium property. When the medium structure evolves over time, the parameter can change. The CTRW model leads to the subdiffusion equation with the time fractional Caputo derivative of order $\alpha$ when the subdiffusion parameter $\alpha$ is assumed to be constant. In practice, these assumptions are met in a homogeneous medium which structure does not change with time. If these assumptions are not met, models with distributed subdiffusion parameter have been used \cite{el}, where superstatistics approach is applied; subdiffusion can be accelerated or delayed depending on the distribution of $\alpha$ \cite{cgs}. In other models subdiffusion equations with a fractional time derivative of the order depending on time and/or on a spatial variable have been used \cite{chen,roth,awad,yzw}. Ultraslow diffusion with parameter evolving in time was considered in Ref. \cite{lwc}. 
If a structure of a diffusing medium changes substantially, the type of diffusion may also change. An example of a process in which the medium structure can be changed is diffusion of an antibiotic in a bacterial biofilm. A biofilm has a gel--like structure and subdiffusion of an antibiotic is expected \cite{km}. Bacteria have different defence mechanisms against the effects of an antibiotic \cite{aot}. One of them is an increasing compaction of the biofilm, which changes the biofilm structure and leads to hindering diffusion of the antibiotic.

In some models of anomalous diffusion different fractional derivatives, not equivalent to each other, have been involved in the diffusion equation. The list of the references regarding this issue is long enough, see for example \cite{ly,skmg}. The examples of anomalous diffusion equations are Erdelyi--Kober fractional diffusion equation \cite{gp}, diffusion equations with Antagana--Baleanu--Caputo and Antagana--Baleanu--Riemann--Liouville fractional derivatives, in which the Mittag--Leffler function is involved in the kernel of fractional derivative operator \cite{yzsz}, a Wiman type \cite{liang} and a Prabhakar--type fractional diffusion equation \cite{skmg,sw2018} in which a kernel of integral operator with respect to time is expressed in terms of the three--parameter Mittag--Leffler function, and Cattaneo--Hristov diffusion equation with Caputo--Fabrizio fractional derivative \cite{hristov}, see also Ref. \cite{fz} and the references cited therein. 
A further generalization of fractional derivatives are fractional $g$--derivatives with respect to another functions $g$ \cite{abd}. We mention that these derivatives are defined frequently for the function denoted as $\psi$ and the derivatives are called $\psi$-fractional derivatives with respect to another function. However, in the analysis of anomalous diffusion processes, the letter $\psi$ is commonly used to denote the distribution of the waiting time for a particle to jump. Therefore, in this paper we denote the derivative with the letter $g$. These derivatives significantly expand the possibilities of defining new diffusion equations. For example, the $g$--Caputo derivative with respect to time \cite{sz} and to spatial variable \cite{boh} have been involved in anomalous diffusion equations. 

Subdiffusion equation with the Caputo or Riemann-Liouville fractional derivative can be solved by means of the Laplace transform method. However, solving some other fractional diffusion equations can require special methods. For example, the solution of the fractional Hilfer-Prabhakar and Cattaneo--Hristov diffusion equation can be obtained using the Elzaki transform \cite{skmg}. Frequently, numerical methods of solving subdiffusion equations with a variable subdiffusion parameter \cite{chen,yzw,wwl} and for equations containing a fractional $g$-derivative \cite{yps} have been used.

The aim of our study is to show an application of the subdiffusion equation with $g$--Caputo fractional derivative to describe subdiffusion in a medium having a structure evolving over time. In the following, we call the subdiffusion equation with fractional $g$--Caputo time derivative as the $g$--subdiffusion equation, and the process described by this equation as $g$--subdiffusion. 
The $g$--subdiffusion process is defined by both: the parameter $\alpha$ and a function $g$, where

(i) $\alpha$ is a subdiffusion parameter for the process taking place in an initial time interval,

(ii) $g$ controls the rest of the process. 

We also show a method of solving the $g$--subdiffusion equation. In particular, we show that this equation describes subdiffusion in a system in which the type of diffusion changes from ``ordinary'' subdiffusion with fixed $\alpha$ to ultraslow diffusion. The Green's function for the process is also found. One of the properties characterizing subdiffusion is the aging process. In general, the aging means that the process is not invariant with the translation in the time domain. ``Standard'' aging of subdiffusion with fixed $\alpha$ is due to a heavy tail of distribution of time which is needed to take a particle step \cite{ks,mjcb,barkai2003,schulz2014,barkaiprl}. For ultraslow diffusion the tail is super-heavy \cite{ckm2017}. We show that the aging of the $g$-subdiffusion process is a combination of ``standard'' subdiffusive aging and an additional aging process described by the function $g$, the latter may be due to changes in the medium structure.

The organization of the paper is as follows. In Sec. \ref{secII} we show a standard Laplace transform method for solving "ordinary" subdiffusion equations with the fractional Caputo time derivative. In Sec. \ref{secIII} we present the method of solving the $g$--subdiffusion equation. In this method the Laplace transform with respect to the function $g$ is used. As an example, we derive the Green's function for a homogeneous system. In Sec. \ref{secIV} we show that the appropriate choice of the $g$ function provides the equation describing the transition process from "ordinary" subdiffusion to ultraslow diffusion. The aging process of $g$--subdiffusion is considered in Sec \ref{secV}. We define the relative aging coefficient $\rho_{\alpha,g}$ of the $g$--subdiffusion process in relation to subdiffusion with a fixed parameter $\alpha$. The final remarks and conclusions are in Sec. \ref{secVI}.

\section{Subdiffusion equation with ``ordinary'' Caputo fractional derivative\label{secII}}

We show how to obtain the Green function for the subdiffusion equation with the ordinary Caputo derivative. Although the results are well known, we present them in some detail as they are the basis for the solving method of $g$-subdiffusion equation. 

The fractional subdiffusion equation with ``ordinary'' Caputo derivative of the order $\alpha\in(0,1)$ is
\begin{equation}\label{eqII1}
\frac{^C \partial^{\alpha} P(x,t|x_0)}{\partial t^\alpha}=D\frac{\partial^2 P(x,t|x_0)}{\partial x^2},
\end{equation}
where the Caputo fractional derivative is defined for $0<\alpha<1$ as
\begin{equation}\label{eqII2}
\frac{^Cd^{\alpha} f(t)}{dt^\alpha}=\frac{1}{\Gamma(1-\alpha)}\int_0^t (t-t')^{-\alpha}f'(t')dt',
\end{equation}
$\alpha$ is a subdiffusion parameter and $D$ is a generalized diffusion coefficient.
To solve the equation the Laplace transform method can be used, the Laplace transform is defined as
\begin{equation}\label{eqII3}
\mathcal{L}[f(t)](s)=\int_0^\infty {\rm e}^{-st}f(t)dt.
\end{equation}
Due to the relation
\begin{equation}\label{eqII4}
\mathcal{L}\left[\frac{^C d^\alpha f(t)}{dt^\alpha}\right](s)=s^\alpha\mathcal{L}[f(t)](s)-s^{\alpha-1}f(0),
\end{equation}
where $0<\alpha\leq 1$, we get
\begin{eqnarray}\label{eqII5}
s^\alpha\mathcal{L}[P(x,t|x_0)](s)-s^{\alpha-1}P(x,0|x_0)\\
=D\frac{\partial^2\mathcal{L}[P(x,t|x_0)](s)}{\partial x^2}.\nonumber
\end{eqnarray}

The Green's function $P(x,t|x_0)$ is the solution to subdiffusion equation for the initial condition 
\begin{equation}\label{eqII6}
P(x,0|x_0)=\delta(x-x_0),
\end{equation}
where $\delta$ is the delta--Dirac function. For unbounded system this function vanishes at $\pm\infty$, the boundary conditions are
\begin{equation}\label{eqII7}
P(-\infty,t|x_0)=P(\infty,t|x_0)=0.
\end{equation}
To solve Eq. (\ref{eqII5}) with the initial condition Eq. (\ref{eqII6}) the standard Fourier transform method can be used. In terms of the Laplace transform the Green's function for the above boundary conditions is
\begin{equation}\label{eqII8}
\mathcal{L}[P(x,t|x_0)](s)=\frac{1}{2\sqrt{D}s^{1-\alpha/2}}\;{\rm e}^{-\frac{s^{\alpha/2}}{\sqrt{D}}|x-x_0|},
\end{equation}
this equation has already been derived in Ref. \cite{barkai2000}.
Using the formula \cite{tk2004}
\begin{eqnarray}\label{eqII9}
\mathcal{L}^{-1}[s^\nu {\rm e}^{-as^\beta}](t)=\frac{1}{t^{1+\nu}}\sum_{k=0}^\infty \frac{1}{k!\Gamma(-\nu-\beta k)}\left(-\frac{a}{t^\beta}\right)^k\\ \equiv f_{\nu,\beta}(t:a),\nonumber
\end{eqnarray}
where $a,\beta>0$, $\Gamma$ is the Gamma-Euler function, we obtain
\begin{equation}\label{eqII10}
P(x,t|x_0)=\frac{1}{2\sqrt{D}}f_{-1+\alpha/2,\alpha/2}\left(t;\frac{|x-x_0|}{\sqrt{D}}\right).
\end{equation}
The function $f$ in Eq. (\ref{eqII10}) is the Mainardi function which is the special case of the Wright function and the H-Fox function, the Mainardi function often appears in solutions to subfiffusion equation \cite{mmp}. We mention that the inverse Laplace transform of Eq. (\ref{eqII8}), Eq. (\ref{eqII10}), can be represented by the inverse one--sided Levy stable density \cite{barkai2001,barkai2002}, see also \cite{wang}.

Since the mean particle position does not change in time and equals $x_0$, the time evolution of the Mean Square Displacement (MSD) $\sigma^2$ of a particle is calculated using the formula 
\begin{equation}\label{eqII11}
\sigma^2(t)=\int_{-\infty}^\infty (x-x_0)^2 P(x,t|x_0)dx.
\end{equation}
From Eqs. (\ref{eqII8}) and (\ref{eqII11}) we get
\begin{equation}\label{eqII12}
\mathcal{L}[\sigma^2(t)](s)=\frac{2D}{s^{1+\alpha}}.
\end{equation}
Using the formula $\mathcal{L}^{-1}[1/s^{1+\nu}]=t^\nu/\Gamma(1+\nu)$, $\nu>-1$, we obtain 
\begin{equation}\label{eqII13}
\sigma^2(t)=\frac{2Dt^\alpha}{\Gamma(1+\alpha)}.
\end{equation}

\section{Subdiffusion equation with g-Caputo fractional derivative\label{secIII}}

We assume that the function $g$, defined for $t\geq 0$, fulfils the conditions $g(0)=0$, $g(\infty)=\infty$, and $g'(t)>0$ for $t>0$.
The $g$-Caputo fractional derivative $^Cd^{\alpha}_g \tilde{f}(t)/dt^\alpha$ of the order $\alpha$ with respect to the function $g$ is defined for $0<\alpha<1$ as 
\begin{equation}\label{eqIII1}
\frac{^Cd^{\alpha}_g \tilde{f}(t)}{dt^\alpha}=\frac{1}{\Gamma(1-\alpha)}\int_0^t (g(t)-g(t'))^{-\alpha}\tilde{f}'(t')dt'.
\end{equation}
The values of function $g$ are given in a time unit. When $g(t)=t$, the $g$-Caputo fractional derivative takes the form of the ``ordinary'' Caputo derivative (\ref{eqII2}).

The $g$-subdiffusion equation reads 
\begin{equation}\label{eqIII2}
\frac{^C \partial^{\alpha}_g \tilde{P}(x,t|x_0)}{\partial t^\alpha}=D\frac{\partial^2 \tilde{P}(x,t|x_0)}{\partial x^2},
\end{equation}
throughout this paper we denote the functions related to the $g$--subdiffusion process described by Eq. (\ref{eqIII2}) with tilde. 
To solve Eq. (\ref{eqIII2}) it is convenient to use the $g$-Laplace transform \cite{jarad} 
\begin{equation}\label{eqIII3}
\mathcal{L}_g[\tilde{f}(t)](s)=\int_0^\infty {\rm e}^{-s g(t)}\tilde{f}(t)g'(t)dt.
\end{equation}
This transform has the following property that makes the procedure for solving Eq. (\ref{eqIII2}) similar to the procedure for solving Eq. (\ref{eqII1}) using the "ordinary" Laplace transform
\begin{equation}\label{eqIII4}
\mathcal{L}_g\left[\frac{^Cd^{\alpha}_g}{dt^\alpha}\tilde{f}(t)\right](s)=s^\alpha\mathcal{L}_g[\tilde{f}(t)](s)-s^{\alpha-1}\tilde{f}(0).
\end{equation}
Both transforms are related to each other by the following relation
\begin{equation}\label{eqIII5}
\mathcal{L}_g[\tilde{f}(t)](s)=\mathcal{L}[\tilde{f}(g^{-1}(t))](s).
\end{equation}
Eq. (\ref{eqIII5}) and the Lerch's uniqueness of the inverse Laplace transform theorem provide the following rule
\begin{equation}\label{eqIII6}
\mathcal{L}_g[\tilde{f}(t)](s)=\mathcal{L}[f(t)](s)\Leftrightarrow \tilde{f}(t)=f(g(t)).
\end{equation}
The above relation is the basis of the method of solving the $g$--subdiffusion equation.

Due to Eq. (\ref{eqIII4}), in terms of the $g$--Laplace transform the $g$--subdiffusion equation is
\begin{eqnarray}\label{eqIII7}
s^\alpha\mathcal{L}_g[\tilde{P}(x,t|x_0)](s)-s^{\alpha-1}\tilde{P}(x,0|x_0)\\
=D\frac{\partial^2\mathcal{L}_g[P(x,t|x_0)](s)}{\partial x^2}.\nonumber
\end{eqnarray}

The structure of Eq. (\ref{eqIII7}) as a differential equation with respect to $x$ variable is the same as the structure of Eq. (\ref{eqII5}). The solution to Eq. (\ref{eqIII7}) for the boundary conditions $\tilde{P}(-\infty,t|x_0)=\tilde{P}(\infty,t|x_0)=0$ and the initial condition
\begin{equation}\label{eqIII8}
\tilde{P}(x,0|x_0)=\delta(x-x_0)
\end{equation}
is
\begin{equation}\label{eqIII9}
\mathcal{L}_g[\tilde{P}(x,t|x_0)](s)=\frac{1}{2\sqrt{D}s^{1-\alpha/2}}\;{\rm e}^{-\frac{s^{\alpha/2}}{\sqrt{D}}|x-x_0|}.
\end{equation}
From Eqs. (\ref{eqII8}) and (\ref{eqIII9}) we obtain
\begin{equation}\label{eqIII10}
\mathcal{L}_g[\tilde{P}(x,t|x_0)](s)=\mathcal{L}[P(x,t|x_0)](s).
\end{equation}
Due to the relation Eq. (\ref{eqIII6}) we have
\begin{equation}\label{eqIII11}
\tilde{P}(x,t|x_0)=P(x,g(t)|x_0),
\end{equation}
and from Eq. (\ref{eqII10}) we get
\begin{equation}\label{eqIII12}
\tilde{P}(x,t|x_0)=\frac{1}{2\sqrt{D}}f_{-1+\alpha/2,\alpha/2}\left(g(t);\frac{|x-x_0|}{\sqrt{D}}\right).
\end{equation}
The time evolution of Mean Square Displacement (MSD) $\tilde{\sigma}^2(t)=\int_{-\infty}^\infty (x-x_0)^2 \tilde{P}(x,t|x_0)dx$ reads
\begin{equation}\label{eqIII13}
\tilde{\sigma}^2(t)=\sigma^2(g(t))=\frac{2D(g(t))^\alpha}{\Gamma(1+\alpha)}.
\end{equation}

\begin{figure}[htb]
\centering{%
\includegraphics[scale=0.4]{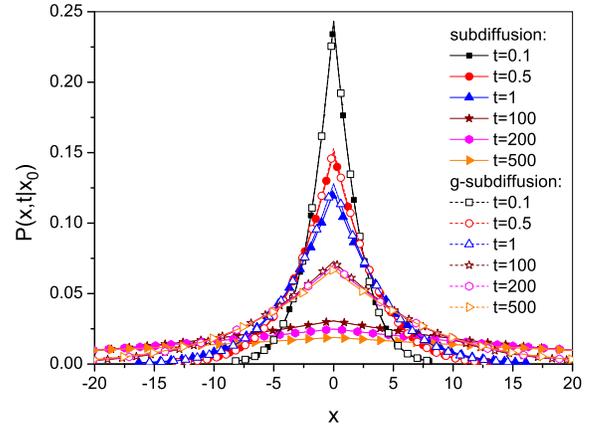}}
\caption{Solutions to the fractional subdiffusion equation Eq. (\ref{eqII10}) (solid lines with filled symbols) and to $g$-subdiffusion equation Eq. (\ref{eqIII12}) with $g$ given by Eq. (\ref{eqIV5}) with $A=2$, $B=1$, $C=0.6$, and $\gamma=2$ (dashed lines with open symbols) for times given in the legend, for both cases the parameters are $\alpha=0.6$, $D=10$, and $x_0=0$, all quantities are given in arbitrarily chosen units.}
\label{fig1}
\end{figure}

\begin{figure}[htb]
\centering{%
\includegraphics[scale=0.4]{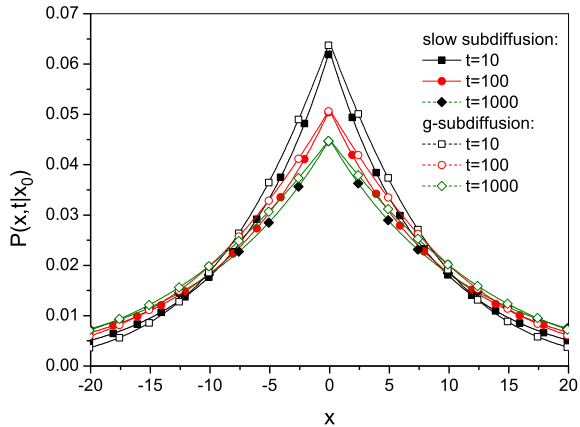}}
\caption{Solutions to the ultraslow diffusion equation Eq. (\ref{eqIV3}) (solid lines with filled symbols) for $\alpha=0.6$ and $D_u=15$, and to $g$-subdiffusion equation Eq. (\ref{eqIII12}) (dashed lines with open symbols) with $g$ given by Eq. (\ref{eqIV5}) for times given in the legend, the other parameters are the same as in Fig. \ref{fig1}.}
\label{fig2}
\end{figure}

\begin{figure}[htb]
\centering{%
\includegraphics[scale=0.4]{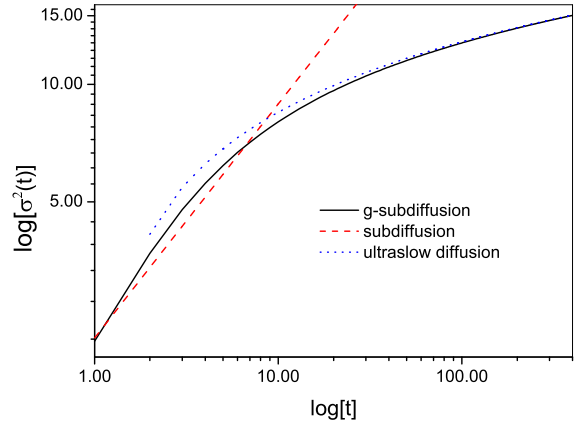}}
\caption{Time evolution of MSD in the logarithmic scale for subdiffusion Eq. (\ref{eqII13}), $g$--subdiffusion Eq. (\ref{eqIII13}), and ultraslow diffusion Eq. (\ref{eqIV4}) for the parameters given in Figs. \ref{fig1} and \ref{fig2}.}
\label{fig3}
\end{figure}

\section{From subdiffusion to ultraslow diffusion\label{secIV}}

When analyzing ultraslow diffusion process, where functions such as logarithm occur, the time variable should be expressed in dimensionless units as $t/\tau_0$ where $\tau_0$ is a parameter given in a time unit. For the sake of simplicity, in the following we assume $\tau_0=1$. Ultraslow diffusion may be defined as a diffusion process in which $\sigma^2(t)\sim \eta(t)$ for $t\rightarrow\infty$, where $\eta$ is a slowly varying function. We consider the following ultraslow diffusion equation 
\begin{eqnarray}\label{eqIV1}
	\frac{1}{\Gamma(\alpha)}\int_0^t \mu(t-t',\alpha-1)\frac{\partial P(x,t';x_0)}{\partial t'}dt'\\
	=D_u\frac{\partial^2 P(x,t;x_0)}{\partial x^2}\;,\nonumber
\end{eqnarray}
$\alpha>0$, where
\begin{equation}\label{eqIV2}
\mu(t,\beta)=\int_0^\infty d\zeta \frac{t^\zeta \zeta^\beta}{\Gamma(1+\zeta)},
\end{equation}
$\beta>-1$, is the Volterra--type function \cite{mainardi,ob}, $D_u$ is a ultraslow diffusion coefficient. In the long time limit the Green's function for Eq. (\ref{eqIV1}) is
\begin{equation}\label{eqIV3}
P(x,t|x_0)=\frac{1}{2\sqrt{D_u{\rm ln}^\alpha t}}{\rm e}^{-\frac{|x-x_0|}{\sqrt{D_u{\rm ln}^\alpha t}}},
\end{equation}
the derivation of Eq. (\ref{eqIV3}) is shown in Appendix.
Eq. (\ref{eqIV3}) provides in the long time limit
\begin{equation}\label{eqIV4}
\sigma^2(t)=2D_u{\rm ln}^\alpha t.
\end{equation}
Subdiffusion of a single particle is described by Eq. (\ref{eqII1}), in this case we have $\sigma^2(t)\sim t^\alpha$. 

We use the $g$-subdiffusion equation to describe a transition process from subdiffusion to ultraslow diffusion. We assume
\begin{equation}\label{eqIV5}
g(t)=\frac{t}{1+Bt}\left[1+A\;{\rm ln}\left(1+Ct^\gamma\right)\right].
\end{equation}
The asymptotic form of the function Eq. (\ref{eqIV5}) is
\begin{eqnarray}\label{eqIV6}
g(t)=\left\{
\begin{array}{c}
t,\;t\rightarrow 0,\\
\frac{A\gamma}{B}\;{\rm ln}\;t,\;t\rightarrow\infty.
\end{array}
\right.
\end{eqnarray}
From Eqs. (\ref{eqIII13}) and (\ref{eqIV6}) we get 
\begin{eqnarray}\label{eqIV7}
\tilde{\sigma}^2 (t)\sim\left\{
\begin{array}{c}
t^\alpha,\;t\rightarrow 0,\\
{\rm ln}^\alpha\;t,\;t\rightarrow\infty.
\end{array}
\right.
\end{eqnarray}

Thus, $g$-subdiffusion equation transforms continuously subdiffusion (at small times) into ultraslow diffusion (at long times). For "moderate" times these processes are mixed. The transition from subdiffusion to ultraslow diffusion is illustrated in Figs \ref{fig1}--\ref{fig3}, all quantities are given in arbitrarily chosen units. In Fig. \ref{fig1} the Green's functions Eq. (\ref{eqII10}) for subdiffusion equation are compared to the Green's functions Eq. (\ref{eqIII12}) for $g$-subdiffusion equation with $g$ given by Eq. (\ref{eqIV5}). For small times good coincidence of these functions is observed. For long times, the Green's functions for $g$-subdiffusion equation cannot be approximated by the Green's functions for subdiffusion equation. In this case the scatter of the plots of Green's function for $g$-subdiffusion equation is much smaller than the scatter of the Green's functions for the ``ordinary'' subdiffusion equation. This fact suggests that for long times the Green's functions Eq. (\ref{eqIII12}) for $g$-subdiffusion equation may describe ultraslow diffusion. A comparison of the solutions to the ultraslow diffusion equation and $g$-subdiffusion equation is shown in Fig \ref{fig2}. This plot shows that solutions to the $g$-subdiffusion equation Eq. (\ref{eqIII12}) can be well approximated by solutions of the ultraslow diffusion equation Eq. (\ref{eqIV3}) for long times. The transition from subdiffusion to ultraslow diffusion is shown in Fig. \ref{fig3} in which time evolution of MSD for three processes is presented.

\section{Aging property of g--subdiffusion process\label{secV}}

One of the aging process features is that the average number of particle jumps in the time interval $(t_a,t_a+\Delta t)$ depends not only on the length of the interval $\Delta t$, but also on the time $t_a$. For subdiffusion described by the equation with "ordinary" Caputo derivative Eq. (\ref{eqII1}) a diffusive medium does not change its properties with time. In this case the aging process is generated by a heavy--tailed distribution of time which is needed for a particle to jump. 

The mean number of jumps in the time interval $(0,t)$, $\left\langle n(t)\right\rangle$, is related to the MSD as follows \cite{ks}
\begin{equation}\label{eqV1}
\sigma^2(t)=l^2\left\langle n(t)\right\rangle,
\end{equation}
where $l^2$ is the variation of the length of a particle jump. From Eqs. (\ref{eqII13}) and (\ref{eqV1}) we get
\begin{equation}\label{eqV2}
\left\langle n(t)\right\rangle=\kappa t^\alpha ,
\end{equation}
where $\kappa=2D/l^2\Gamma(1+\alpha)$. The mean number of particle jumps in the time interval $(t_a,t_a+\Delta t)$ is
\begin{equation}\label{eqV3}
\left\langle n(t_a,\Delta t)\right\rangle=\left\langle n(t_a+\Delta t)\right\rangle-\left\langle n(t_a)\right\rangle.
\end{equation}
The above function is usually considered for two extreme cases of $\Delta t\ll t_a$ and $\Delta t\gg t_a$.
From Eqs. (\ref{eqV2}) and (\ref{eqV3}) we get
\begin{equation}\label{eqV4}
\left\langle n(t_a,\Delta t)\right\rangle=\left\{
\begin{array}{c}
\kappa \alpha t_a^{\alpha-1}\Delta t,\;\Delta t\ll t_a,\\
   \\
\kappa (\Delta t)^\alpha,\;\Delta t\gg t_a.
\end{array}
\right.
\end{equation}
Eq. (\ref{eqV4}) has already been derived in Ref. \cite{schulz2014}, see also \cite{barkaiprl}.
For ultraslow diffusion we obtain $\left\langle n_u\right\rangle$ from Eqs. (\ref{eqIV4}), (\ref{eqV1}), and (\ref{eqV3})
\begin{equation}\label{eqV5}
\left\langle n_u(t_a,\Delta t)\right\rangle=\left\{
\begin{array}{c}
\frac{\kappa_u}{t_a {\rm ln}^{1-\alpha}(t_a)}\Delta t,\;\Delta t\ll t_a,\\
   \\
\kappa_u {\rm ln}^{\alpha}(\Delta t),\;\Delta t\gg t_a,
\end{array}
\right.
\end{equation}
where $\kappa_u=2D_u/l^2\Gamma(1+\alpha)$.
For the $g$--subdiffusion process we have
\begin{equation}\label{eqV6}
\left\langle \tilde{n}(t)\right\rangle=\kappa g^{\alpha}(t),
\end{equation}
and 
\begin{equation}\label{eqV7}
\left\langle \tilde{n}(t_a,\Delta t)\right\rangle=\left\langle \tilde{n}(t_a+\Delta t)\right\rangle-\left\langle \tilde{n}(t_a)\right\rangle.
\end{equation}
From Eqs. (\ref{eqV6}) and (\ref{eqV7}) we get
\begin{equation}\label{eqV8}
\left\langle \tilde{n}(t_a,\Delta t)\right\rangle=\left\{
\begin{array}{c}
\kappa \alpha g^{\alpha-1}(t_a)g'(t_a)\Delta t,\;\Delta t\ll t_a,\\
   \\
\kappa g^\alpha(\Delta t),\;\Delta t\gg t_a.
\end{array}
\right.
\end{equation}
We consider the aging effect for relatively small $\Delta t$, when $\Delta t\ll t_a$. Let us define the relative aging coefficient $\rho_{g,\alpha}$ for the $g$-subdiffusion process with the parameter $\alpha$,
\begin{equation}\label{eqV9}
\rho_{g,\alpha}(t_a)=\frac{\left\langle \tilde{n}(t_a,\Delta t)\right\rangle}{\left\langle n(t_a,\Delta t)\right\rangle}
=\frac{g^{\alpha-1}(t_a)g'(t_a)}{t_a^{\alpha-1}}.
\end{equation}
The coefficient $\rho_{g,\alpha}$ shows the relation of the aging effect generated in the $g$--subdiffusion process to the aging effect in the ``ordinary'' subdiffusion process with a fixed $\alpha$. For the process considered in Sec. \ref{secIV} we have
\begin{eqnarray}\label{eqV10}
\rho_{g,\alpha}(t_a)=\frac{\left[1+A\;{\rm ln}(1+Ct_a^\gamma)\right]^{\alpha-1}}{(1+Bt_a)^\alpha}\\
\times \left[\frac{1+A\;{\rm ln}(1+Ct_a^\gamma)}{1+Bt_a}+\frac{AC\gamma t_a^\gamma}{1+Ct_a^\gamma}\right].\nonumber
\end{eqnarray}

\begin{figure}[htb]
\centering{%
\includegraphics[scale=0.4]{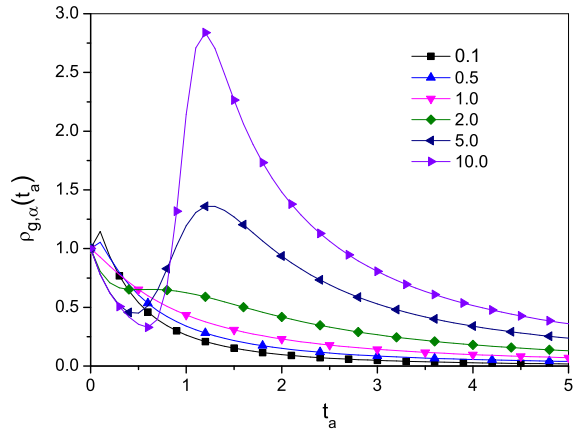}}
\caption{Plots of the function Eq. (\ref{eqV10}) for different $\gamma$ given in the legend, here $\alpha=0.6$, the other parameters are the same as in Fig. \ref{fig1}.}
\label{fig4}
\end{figure}

\begin{figure}[htb]
\centering{%
\includegraphics[scale=0.4]{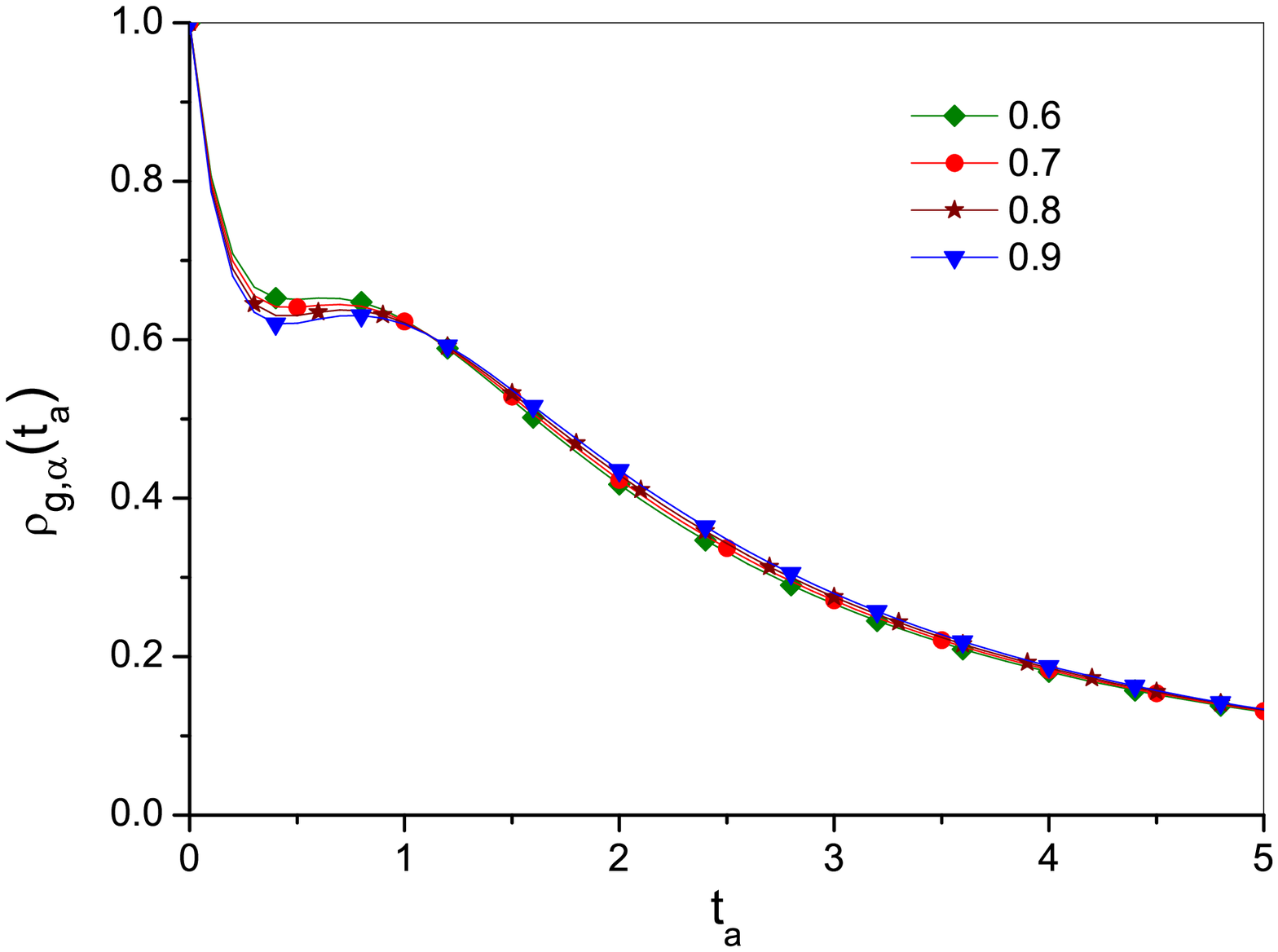}}
\caption{Plots of the function Eq. (\ref{eqV10}) for different $\alpha$ given in the legend, here $\gamma=2.0$, the other parameters are the same as in Fig. \ref{fig1}.}
\label{fig5}
\end{figure}

\begin{figure}[htb]
\centering{%
\includegraphics[scale=0.4]{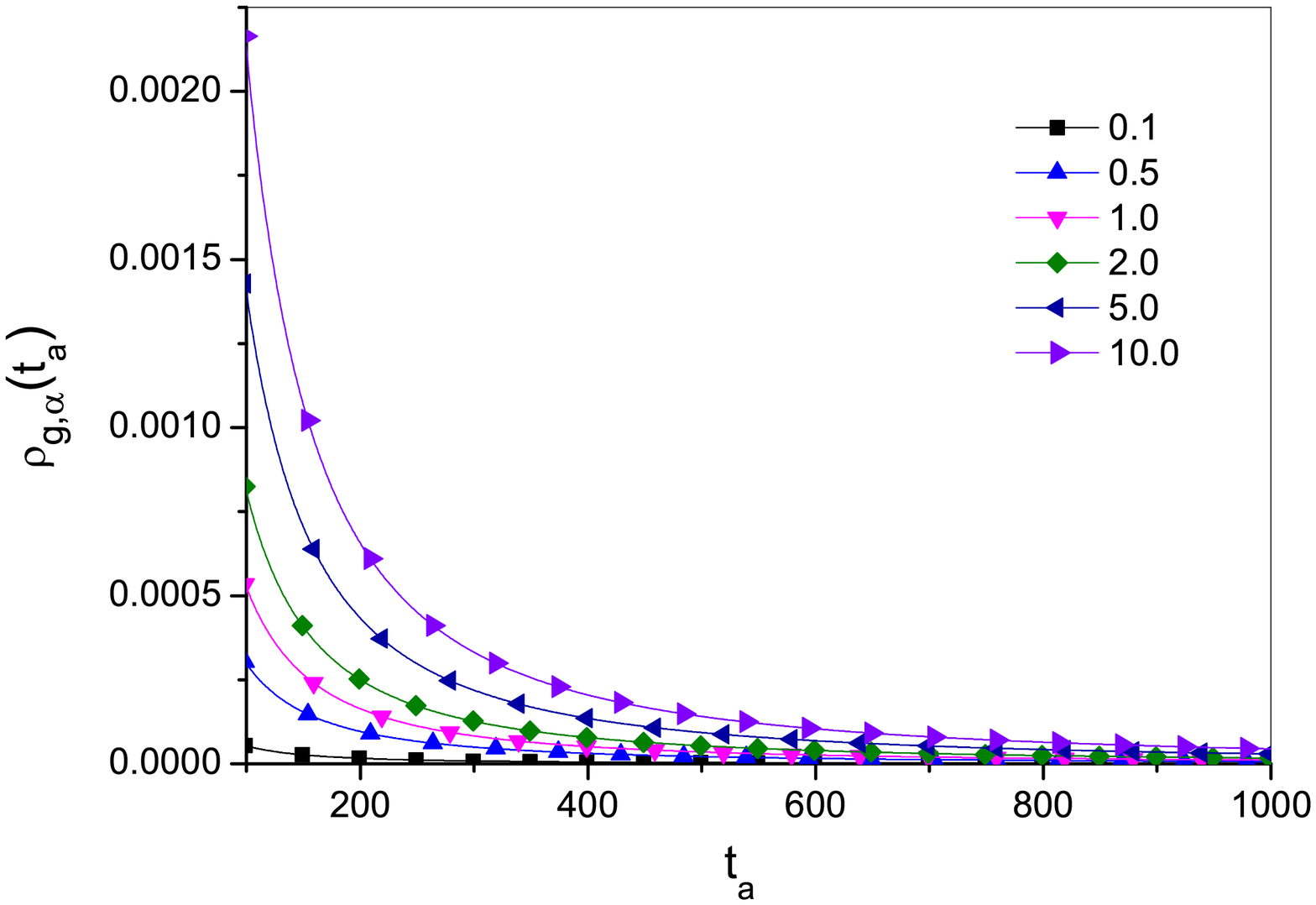}}
\caption{Plots of the function Eq. (\ref{eqV10}) for $t\in [100,1000]$, the description is as in Fig. \ref{fig4}}. 
\label{fig6}
\end{figure}

\begin{figure}[htb]
\centering{%
\includegraphics[scale=0.4]{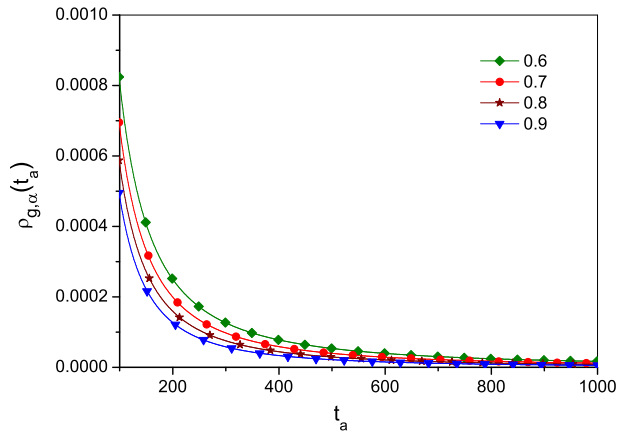}}
\caption{Plots of the function Eq. (\ref{eqV10}) for $t\in [100,1000]$, the description is as in Fig. \ref{fig5}}. 
\label{fig7}
\end{figure}

In the limit of long time, we have $\rho_{\alpha,g}(t_a)\sim 1/t_a^\alpha {\rm ln}^{1-\alpha}(t_a)$. When $t_a\rightarrow\infty$ we have $\rho_{g,\alpha}(t_a)\rightarrow 0$. Based on Figs. \ref{fig4}--\ref{fig7} we briefly consider the influence of two exponents $\alpha$ and $\gamma$ on the function $\rho_{\alpha,g}$.
Figs \ref{fig4}--\ref{fig7} show the dependence of the coefficient $\rho_{g,\alpha}$ on time $t_a$. The plots have been made for different $\gamma$ (Figs. \ref{fig4} and \ref{fig6}) and for different $\alpha$ (Figs. \ref{fig5} and \ref{fig7}). For small $t_a$ the coefficient $\gamma$ apparently affects on the coefficient $\rho_{g,\alpha}$, while the effect of the parameter $\alpha$ is barely noticeable. For relatively long times, the effect of $\alpha$ on $\rho_{\alpha.g}$ is greater than for small times, but it is much smaller than the effect of the exponent $\gamma$. This is because the coefficient $\rho_{\alpha,g}$ describes a change in the aging process of $g$--subdiffusion compared to aging of "ordinary" subdiffusion when $\alpha$ is the same for both processes. Then, the effect of $\alpha$ on $\rho_{\alpha,g}$ is relatively small.

\section{Final remarks\label{secVI}}

The subdiffusion process in which the subdiffusion parameter $\alpha$ as well as a type of diffusion may change in time can be described by an equation containing the fractional Caputo time derivative with respect to another function $g$. This equation has been called the $g$-subdiffusion equation and it describes a $g$--subdiffusion process. The process is defined by the parameter $\alpha$ and the function $g$. The final remarks and conclusions are as follows.

(i) At some initial time interval $g$--subdiffusion is described by a fractional subdiffusion equation with a fixed parameter $\alpha$ which corresponds to the $g$--subdiffusion equation for $g(t)=t$. Thus, the general form of the function $g$ is
\begin{equation}\label{eqVI1}
g(t)=t+h(t),
\end{equation}
where the function $h$ fulfils the conditions $h(0)=0$, $h'(t)>-1$ and $h(t)>-t$ for $t>0$. For $g$ given by Eq. (\ref{eqIV5}) we have $h(t)=t[A\;{\rm ln}(1+Ct^\gamma)-Bt]/(1+Bt)$, in the limit of long time there is $h(t)<0$. The subdiffusion parameter $\alpha$ depends on the structure of the medium. If the structure evolves over time such that it affects subdiffusion, then $h(t)\neq 0$. If the change in the properties of the medium leads to an additional difficulty in subdiffusion, then $h(t)<0$. When the change of medium structure facilitates subdiffusion (e.g. the density of the medium is reduced), then $h(t)>0$. We mention that normal diffusion can be considered here as a special case of subdiffusion for which $\alpha=1$; then the subdiffusion equation Eq. (\ref{eqII1}) takes the form of the normal diffusion equation. 

(ii) For the function 
\begin{equation}\label{eqVI2}
g(t)=t+D_\beta t^{\beta/\alpha},
\end{equation}
where $\beta>\alpha$, we get the relation $\tilde{\sigma}^2(t)\sim t^\beta$ in the long time limit. For $\beta>1$ we get the relation characteristic for superdiffusion. Apparently, it is possible to apply the $g$-subdiffusion equation to describe superdiffusion. However, within the CTRW model superdiffusion is created by anomalously long particle jumps which can be done with relatively high probabilities. The probability density of the particle jump length has a heavy tail. This leads to the superdiffusion equation with the fractional derivative with respect to a spatial variable. This is not the case considered within the $g$-subdiffusion model in which a type of diffusion is defined by the function $g(t)$. Therefore, in our opinion, the problem of whether transition from subdiffusion to superdiffusion can be described by the $g$-subdiffusion equation is still open.

(iii) Replacing in the subdiffusion equation the "ordinary" fractional Caputo derivative by the $g$--Caputo derivative provides a rescaling of the time variable. In general, the change of the time scale in the particle random walk model can lead to subdiffusion \cite{hilfer}. Changing time scale can be made by means of subordinated method when two stochastic processes are entangled with each other, one of them randomly sets the operating time \cite{ks,sokol,sw,feller}. Examples of processes that lead to a rescaling of a diffusion are passages through the layered media \cite{carr}, local rules for transporting molecules which imply that each step of a molecule is a multi-step process \cite{ncl}, anomalous diffusion in an expanding medium \cite{vay}, diffusing diffusivities where the diffusion coefficient evolves over time \cite{csm}, aging phenomenon \cite{mjcb}, and positional resetting process \cite{bcm}. In the $g$--subdiffusion process time is rescaled by the deterministic function $g$. 

(iv) The change of times scale influences on the aging process. As an example, we have considered aging process which is manifested by time evolution of the average number of a diffusing particle jumps doing in a relatively short period of time. This function depends explicitly on $t_a$. The function $\rho_{g,\alpha}$ can be treated as a ``measure'' how far is the $g$--subdiffusion aging process from aging of ``ordinary'' subdiffusion. We have shown that the transition from subdiffusion to ultraslow diffusion creates an "additional" aging process which provides $\left\langle \tilde{n}(t_a,\Delta t) \right\rangle\ll \left\langle n(t_a,\Delta t) \right\rangle$ when $t_a\rightarrow\infty$.

(v) We have shown the procedure of solving the $g$-subdiffusion equation. The procedure consists of two stages: 

(a) the subdiffusion equation with the ordinary Caputo derivative Eq. (\ref{eqII1}) with a fixed parameter $\alpha$ is firstly solved, 

(b) next, we put $t\rightarrow g(t)$ in the obtained solution. 

We have focused our attention on determining the Green's function for $g$-subdiffusion equation for an unbounded system. Using the methods of images the Green's function $\tilde{P}(x,t|x_0)$ can be derived for a system with fully impermeable walls and/or fully absorbing walls \cite{feller,chandra} as well as with a partially absorbing wall \cite{tk2015}. If particles diffuse independently of each other, the concentration of particles $\tilde{C}(x,t)$ being a solution to the $g$--subdiffusion equation can be calculated for any initial concentration $\tilde{C}(x,0)$ using the formula $\tilde{C}(x,t)=\int_\Omega \tilde{P}(x,t|x_0)\tilde{C}(x_0,0)dx_0$, where $\Omega$ is a particle position domain. 

Let $C(x,t)$ be the solution to subdiffuion equation Eq. (\ref{eqII1}) with initial and boundary conditions as for the $g$--subdiffusion equation with the same $\alpha$. We assume that the boundary conditions do not depend explicitly on time. Then, we get $\mathcal{L}_g[\tilde{C}(x,t)]=\mathcal{L}[C(x,t)]$. Due to Eq. (\ref{eqIII6}) we get $\tilde{C}(x,t)=C(x,g(t))$. Thus, the procedure for solving the $g$--subdiffusion equation can be quite widely used.
 
(vi) Concluding, if the medium structure significantly changes over time, the diffusion process can be described by the $g$-subdiffusion equation. The function $g$ depends on a time evolution of the medium structure. In general, $g$--subdiffusion equation can describe a process in which the subdiffusion parameter changes with time. If the changes are very strong, we have a process in which the type of diffusion changes continuously. As we have mentioned in Sec. \ref{secI}, such processes may occur in antibiotic diffusion in a bacterial biofilm. The time evolution of the medium structure provides changes in the aging process. The measure of an additional aging effect is expressed by the coefficient $\rho_{\alpha,g}$. The diffusion processes in which the parameter $\alpha$ changes have been described, among others, by subordinated method using Laplace exponent with two indexes \cite{sw,sw2019}, bi--fractional equation \cite{smc}, and bi--exponent distribution of time to take a particle next step \cite{awad,wcd}. In Ref. \cite{trs} diffusion on comb--like structured medium with two annealing mechanisms was studied. One of them, typical for subdiffusion with fixed $\alpha$, is static and created by quenched disorder, the other is created by an annealed disorder mechanism. Processes as the mentioned above can be described by the $g$--subdiffusion equation with appropriate chosen function $g$. 

\section*{Acknowledgments}

The authors thank Eli Barkai for his comments on subdiffusion equations and aging process, and for pointing out some references.

\section*{Appendix. Green's function for Eq. (\ref{eqIV1})\label{secA}}

Since $\mathcal{L}[\mu(t,\alpha-1)](s)=\Gamma(\alpha)/s\;{\ln}^\alpha s$, $\alpha>0$ \cite{ob}, the slow subdiffusion equation Eq. (\ref{eqIV1}) reads in terms of the Laplace transform
\begin{eqnarray}\label{a1}
\frac{v(s)}{s}\left[s\mathcal{L}[P(x,s|x_0)](s)-\delta(x-x_0)\right]\\
=D_s\frac{\partial^2\mathcal{L}[P(x,s|x_0)](s)}{\partial x^2},\nonumber
\end{eqnarray} 
where $v(s)=1/{\rm ln}^\alpha(1/s)$.
The solution to Eq. (\ref{a1}) is
\begin{equation}\label{a2}
\mathcal{L}[P(x,s|x_0)](s)=\frac{\sqrt{v(s)}}{2s\sqrt{D_s}}{\rm e}^{-|x-x_0|\sqrt{\frac{v(s)}{D_s}}}.
\end{equation}
The strong Tauberian theorem states that the relations $\mathcal{L}[\phi(t)](s)\approx\mathcal{R}(s)/s^\rho$ as $s\rightarrow 0$ and $\phi(t)\approx \mathcal{R}(1/t)/\Gamma(\rho)t^{1-\rho}$ as $t\rightarrow\infty$ implies the other under conditions that $\rho>0$, $\mathcal{R}$ is a slowly varying function, and $\phi(t)\geq 0$ is ultimately monotonic function like $t\rightarrow\infty$. Applying this theorem to Eq. (\ref{a2}), we get Eq. (\ref{eqIV3}) in the long time limit.

\end{document}